\documentclass[twocolumn,showpacs,showkeys,preprintnumbers,amsmath,amssymb,prb,a4paper]{revtex4}
\usepackage{graphicx}
\usepackage{dcolumn}
\usepackage{bm}
\usepackage{rotating}
\newcommand{\nf}{NbFe$_{2}$~}
\newcommand{\nfy}{Nb$_{1-y}$Fe$_{2+y}$~}
\newcommand{\et}{~\textit{et al.~}}
\begin{document}
\preprint{APS/123-QED}
\title{Magnetism in \nfy -- composition and magnetic field dependence}
\author{D. Moroni-Klementowicz}
\author{M. Brando}
\email{manuel.brando@cpfs.mpg.de}
\altaffiliation{New address: Max-Plank-Institut f\"ur Chemische Physik fester Stoffe, N\"othnitzer Strasse 40, D-01187 Dresden, Germany.}
\author{C. Albrecht}
\author{W. J. Duncan}
\author{F. M. Grosche}
\altaffiliation{New address: Cavendish Laboratory, University of Cambridge, J J Thomson Avenue, Cambridge CB3 0HE, UK.}
\affiliation{Department of Physics, Royal Holloway, University of London, Egham TW20 0EX, UK.}
\author{D. Gr\"uner}
\author{G. Kreiner}
\affiliation{Max-Plank-Institut f\"ur Chemische Physik fester Stoffe, N\"othnitzer Strasse 40, D-01187 Dresden, Germany.}
\date{Received \today}
\begin{abstract}
We present a systematic study of transport and thermodynamic
properties of the Laves phase system Nb$_{1-y}$Fe$_{2+y}$. Our
measurements confirm that Fe-rich samples, as well as those rich in
Nb (for $\mid y\mid\geq 0.02$), show bulk ferromagnetism at low
temperature. For stoichiometric NbFe$_2$, on the other hand,
magnetization, magnetic susceptibility and magnetoresistance results
point towards spin-density wave (SDW) order, possibly helical, with a
small ordering wavevector $Q \sim 0.05$~\AA$^{-1}$.

Our results suggest that on approaching the stoichiometric
composition from the iron-rich side, ferromagnetism changes into
long-wavelength SDW order. In this scenario, $Q$
changes continuously from 0 to small, finite values at a Lifshitz
point in the phase diagram, which is located near $y=+0.02$.  Further
reducing the Fe content suppresses the SDW transition
temperature, which extrapolates to zero at $y\approx -0.015$.
Around this Fe content magnetic fluctuations dominate the temperature dependence
of the resistivity and of the heat capacity which deviate from their
conventional Fermi liquid forms, inferring the presence of a
quantum critical point. Because the critical point is located between the
SDW phase associated with stoichiometric NbFe$_2$ and the
ferromagnetic order which reemerges for very Nb-rich NbFe$_2$, the
observed temperature dependences could be attributed both to
proximity to SDW order or to ferromagnetism.
\end{abstract}

\pacs{75.30.Fv, 75.30.Kz, 75.40.Cx, 75.47.-m, 75.50.Bb}
\keywords{NbFe$_{2}$, Laves phase, itinerant magnetism, strongly correlated metals, quantum criticality}
\maketitle
\section{\label{sec:introduction}Introduction}
\noindent
Magnetic quantum phase transitions, where the threshold of magnetism
can be reached at low temperature, guide towards new ordering
phenomena in metals.  While examples of quantum phase transitions and
of the associated quantum critical behavior abound in 4\textit{f}-electron
metals, such as the heavy fermion Ce- and Yb-compounds,\cite{stewart2006}
comparatively few transition metal compounds have been studied in detail.
Among these are the nearly or weakly ferromagnetic (FM) materials MnSi,
Ni$_3$Al/Ni$_3$Ga, $\epsilon$-Fe, FeGe and ZrZn$_2$, as well as layered
oxides such as the high-$T_c$ cuprates and the ruthenates.
\cite{pfleiderer2007,niklowitz2005,aguayo2004,pedrazzini2007,smith2008,broun2008,grigera2004}
Outside the oxide family, itinerant antiferromagnetism, or
spin density wave order, is rare in transition metal compounds. The
most thoroughly studied example is chromium and its alloy series with
vanadium, in which signatures of Fermi liquid (FL)\cite{baym1991} breakdown have been
observed at the quantum critical composition
Cr$_{1-x}$V$_{x}$.\cite{yeh2002} Recently, we have reported anomalous temperature
dependences of the heat capacity and of the electrical resistivity in
slightly off-stoichiometric NbFe$_{2}$.\cite{brando2008} Positioned very close
to the threshold of magnetism at ambient pressure and stoichiometric
composition, \nf may provide fresh opportunities in this long-standing problem.

The C14 Laves phase system \nfy exhibits three magnetically ordered
low temperature states within a narrow composition range at ambient
pressure. At slightly off-stoichiometric compositions, both towards
the iron-rich and the niobium-rich side, it has been reported to be
ferromagnetic at low temperature, with composition-dependent, low
transition temperatures $T_c$ of the order of tens of Kelvin.
Moreover, at and very close to the stoichiometric composition,
\nf has been reported to assume antiferromagnetic or spin density
wave (SDW) order below a second transition temperature $T_m \simeq 20 \rm
 ~K$.\cite{shiga1987,yamada1988,yamada1990,crook1995,kurisu1997} The easy accessibility of these ordered
states in a nearly stoichiometric compound presents a number of
interesting opportunities for studies of quantum phase transitions and
quantum criticality in a transition metal compound.
 
Firstly, it may be possible to examine ferromagnetic quantum
criticality by tuning $T_c$ towards absolute zero. This requires
starting with a niobium-rich sample and either changing the
composition or applying hydrostatic pressure. In iron-rich samples,
ferromagnetism transforms into the presumed spin density wave state on
approaching stoichiometry. The precise nature of this
non-ferromagnetic ordered phase in nearly stoichiometric NbFe$_{2}$, and the
associated quantum critical phenomena, represent a second area of
interest.  Finally, the intermediate composition ranges, at which
ferromagnetism is replaced by the non-ferromagnetic order
characteristic of stoichiometric \nf invite detailed investigation.

Past studies of the \nfy system have explored the
composition-temperature phase diagram in polycrystals by magnetic
measurements and by nuclear magnetic resonance.\cite{yamada1988,crook1995}
While in broad agreement, these studies
differ in their classification of the slightly niobium-rich region of
the phase diagram. The former (Ref.~13)
describes the region around $y\approx -0.01$
as paramagnetic down to $2 \rm ~K$, whereas the latter (Ref.~14)
reports a mixed (FM and SDW) phase.
Further microscopic probes have been
M\"o\ss bauer spectroscopy, muon spin relaxation ($\mu$sr) and neutron
scattering studies. While $\mu$sr has shown evidence of static
moments in stoichiometric NbFe$_{2}$, neutron scattering has so far not
revealed any information about the nature of this magnetic order in NbFe$_{2}$.
%
%
Here, we reexamine the magnetic phase diagram of \nfy in
well-characterized high-quality samples. We aim to
address key questions thrown up by earlier studies:
(i) Are both FM phases previously observed truly belonging
to the C14 phase of \nfy or is the ferromagnetism due to the presence
of a FM second phase;\cite{bi1996,zhu2003} (ii) which is then the real
origin of the SDW order; (iii) how does the SDW state grow out of the FM
state; (iv) does the antiferromagnetism disappear in slightly Nb-rich
samples, and, if so, (v) is there a QCP?
%
\section{\label{sec:experimental_details}Experimental details}
\subsection{\label{subsec:samples_preparation}Sample preparation}
\noindent
Most measurements in previous studies,\cite{shiga1987,yamada1988,crook1995}
were carried out on polycrystalline samples of \nfy prepared by
arc-melting, followed by annealing for one week at
$1000\,^{\circ}\mathrm{C}$. X-ray powder diffraction was used
to check for phase purity and to determine
the lattice parameters, from which the composition was estimated.
\begin{figure}
\begin{center}
\includegraphics[width=0.7\columnwidth,angle=-90]{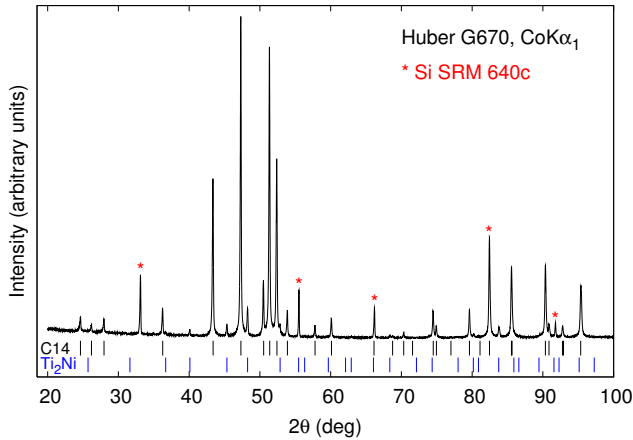}
\end{center}
\caption{(color online). X-ray powder diffractogram of a sample extracted from the upper slice
  of a pellet prepared with method B. Besides reflections of the C14 phase a
  second phase with Ti$_{2}$Ni structure type is clearly observable (possible
  reflection positions are indicated). Reflections of the internal Silicon
  standard are marked with asterisks.}
\label{fig:x-ray}
\end{figure}
Since C14-\nfy crystallizes in a broad homogeneity range from 27.4 to 36.3
at.$\%$ Nb at $1100\,^{\circ}\mathrm{C}$, and because the
magnetic properties are very sensitive to $y$, it is important to
verify not only the structure but also the final composition. We
decided, therefore, to prepare samples in the range from 26 to 40 at.\% Nb,
covering the whole homogeneity range of the C14
phase, with two similar methods (indicated as method A and B in
table~\ref{tab:table1}) accompanied by characterization with X-ray
powder diffraction, metallographic analysis, and wavelength and energy
dispersive X-ray spectroscopy (WDXS and EDXS).
\begin{figure}
\begin{center}
\includegraphics[width=\columnwidth,angle=0]{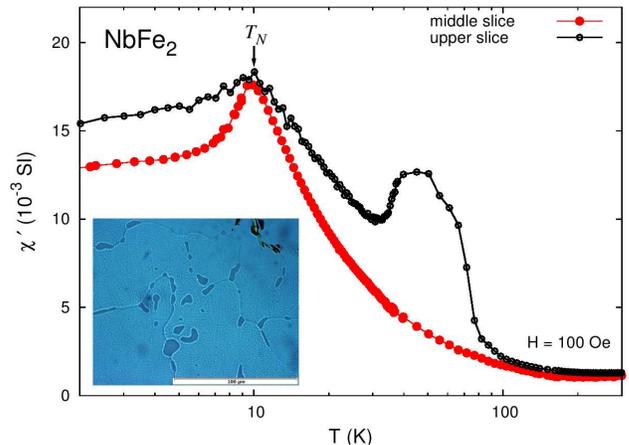}
\end{center}
\caption{(color online). AC-susceptibility of samples cut from the upper and middle
  slice of a pellet prepared with method B (samples with $y=0$ and
  $y=0.003$ in Tab.~\ref{tab:table1}). The measurements performed at
  low magnetic field (100~Oe) allow to detect the presence of FM
  phases ($T_{s}\approx 80$ ~K) in the upper slice sample in addition
  to the expected SDW peak ($T_{N}\approx 10$ ~K). Comparing this
  measurement with the micrograph shown in the inset (the scale length
  is $100~\mu$m), we deduce that the FM signal is caused by the second
  phase present around the \nf grain boundaries.}
\label{fig:sample4b_p4}
\end{figure}
The results are displayed in Table~\ref{tab:table1}. Method A: Pellets
with a mass of $\simeq 2~\rm g$ were prepared by arc melting from pure
niobium (H. C. Starck, granules, $99.9\%$) and iron (Chempur, foil,
$99.997\%$) on a water-cooled copper hearth in an argon atmosphere.  In
order to assure homogeneity, the pellets were turned over and
remelted several times. The pellets were subsequently enclosed in
weld-sealed niobium ampoules which were in turn jacketed in quartz
tubes, and annealed at $1100\,^{\circ}\mathrm{C}$ for three
weeks. Method B: Polycrystals were prepared by melting together the
elements in a radio-frequency induction heating system (in argon
atmosphere) on a water-cooled copper boat, immediately followed by a
short anneal above $1000\,^{\circ}\mathrm{C}$.

X-ray powder diffraction data (Fig.~\ref{fig:x-ray}) have been
collected on a Huber G670 diffractometer equipped with a Ge
monochromator. Co~K$\alpha_1$ radiation was used instead of standard
Cu~K$\alpha_1$ because of the strong absorption of iron. Silicon
(National Institute of Standards \& Technology SRM 640c, $a =
5.43119(1)$\,\AA was used as the internal standard. The lattice
parameters were refined by least-squares fits of the diffraction
angles in the range $20^{\circ} \le 2\theta \le 100^{\circ}$, where
$2\theta$ is the diffraction angle, using the program
PPLP.\cite{gabe1989}
\begin{sidewaystable}
\caption{\label{tab:table1}Results of the phase analysis: The deviation from stoichiometry given by $y$ has been derived from equation~\ref{eq:volume}.}
\begin{ruledtabular}
\begin{tabular}{lrllllllllll}
Nb$_{1-y}$Fe$_{2+y}$ & $y$ & meth. & phase & nom. (at.\%) & WDXS (at.\%) & $a$ (\AA) & $c$ (\AA) & RRR & $T_{N}$(K) & $T_{c,s}$(K)\footnotemark[2] & $\mu_{0}H_{c}$(T)\footnotemark[3]\\
\hline
 Nb$_{0.8}$Fe$_{2.2}$     &    0.2 & (A) & C14 + Fe(Nb) & 26.0 & 27.9(2) & 4.8136(4) & 7.8509(5) & & & &\\
 Nb$_{0.825}$Fe$_{2.175}$ &  0.175 & (A) & C14 + Ti$_{2}$Ni & 28.0 & 28.4(3) & 4.8159(7) & 7.857(1) & & &\\
 Nb$_{0.865}$Fe$_{2.135}$ &  0.135 & (A) & C14\footnotemark[1] & 29.0 & 29.2(2) & 4.8211(5) & 7.8642(7) & & &\\
 Nb$_{0.897}$Fe$_{2.103}$ &  0.103 & (A) & C14\footnotemark[1] & 30.0 & 30.3(3) & 4.8249(6) & 7.872(1) & & &\\
 Nb$_{0.96}$Fe$_{2.04}$   &   0.04 & (A) & C14 & 32.0 & 32.5(2) & 4.8346(5) & 7.8880(8) & & & 72 &\\
 Nb$_{0.992}$Fe$_{2.008}$ &  0.008 & (A) & C14 & 33.3 & 33.2(2) & 4.8419(4) & 7.8978(6) & & 14 &\\
 Nb$_{1.035}$Fe$_{1.965}$ & -0.035 & (A) & C14 & 34.0 & 34.5(3) & 4.8461(6) & 7.9048(9) & & & 14.5 &\\
 Nb$_{1.06}$Fe$_{1.94}$   &  -0.06 & (A) & C14 + Ti$_{2}$Ni & 35.0 & 34.6(1) & 4.8503(5) & 7.9106(7) & & & 30\\
 Nb$_{1.09}$Fe$_{1.91}$   &  -0.09 & (A) & C14 + $\mu$ & 36.0 & 34.8(2) & 4.8556(4) & 7.9175(7) & & &\\
 Nb$_{1.093}$Fe$_{1.907}$ & -0.093 & (A) & C14 + $\mu$ & 38.0 & & 4.8562(6) & 7.918(1) & & &\\
 Nb$_{1.097}$Fe$_{1.903}$ & -0.097 & (A) & C14 + $\mu$ & 40.0 & 35.2(2) & 4.8569(7) & 7.919(2) & & &\\
\hline
Nb$_{1.006}$Fe$_{1.994}$  & -0.006 & (A) & C14 & & 33.3 & 4.8414(6) & 7.8989(1) & 13 & 5 &\\
Nb$_{1.012}$Fe$_{1.988}$  & -0.012 & (A) & C14 & & 33.53 & 4.842(1) & 7.900(2) & & 2 &\\
\hline
NbFe$_{2}$                & 0 & (B) & C14 & 33.3 & 33.3(4) & 4.8401(2) & 7.8963(6) & 9.1 & 10 & & 0.6\\
Nb$_{0.997}$Fe$_{2.003}$  & 0.003 & (B) & C14 + Ti$_{2}$Ni & & & 4.8397(6) & 7.8959(9) & & 10 & 80 &\\
Nb$_{0.993}$Fe$_{2.007}$  & 0.007 & (B) & C14\footnotemark[1] & & & 4.8414(6) & 7.8981(9) & 5.3 & 18 & 6 & 0.15\\
Nb$_{0.985}$Fe$_{2.015}$  & 0.015 & (B) & C14 & & & & & 6 & 32 & 25&\\
\hline
Nb$_{1.01}$Fe$_{1.99}$\footnotemark[4]   & -0.01  & (C) & C14 & & & & & 18 & 2.8 & & 0.2\\
Nb$_{1.022}$Fe$_{1.978}$\footnotemark[4] & -0.022 & (C) & C14 & & & 4.8440(4) & 7.9015(5) & & & 4 &\\
\end{tabular}
\end{ruledtabular}
\footnotetext[1]{Some traces of Ti$_{2}$Ni-structure type phase.}
\footnotetext[2]{$T_{s}$ indicates the FM transition temperature of the second phase with Ti$_{2}$Ni-structure type, while $T_{c}$ refers to the FM state of the C14 Laves phase.}
\footnotetext[3]{Values estimated at 100 m~K}
\footnotetext[4]{Single crystal obtained using Czochralski method.\cite{brando2007,brando2008}}
\end{sidewaystable}
Together with the expected hexagonal C14 Laves phase (space group
$P6_3/mmc$), X-ray powder diffractograms show evidence of a second
phase with Ti$_{2}$Ni structure type (space group $Fd\bar{3}m$, $a
\approx 11.3$\,\AA{}) in almost all investigated samples. The structure
type of the second phase was confirmed for the sample at the composition
Nb$_{1.06}$Fe$_{1.94}$ (see Tab.~\ref{tab:table1}) by electron-back-scatter diffraction.
\begin{figure}
\begin{center}
\includegraphics[width=0.7\columnwidth,angle=-90]{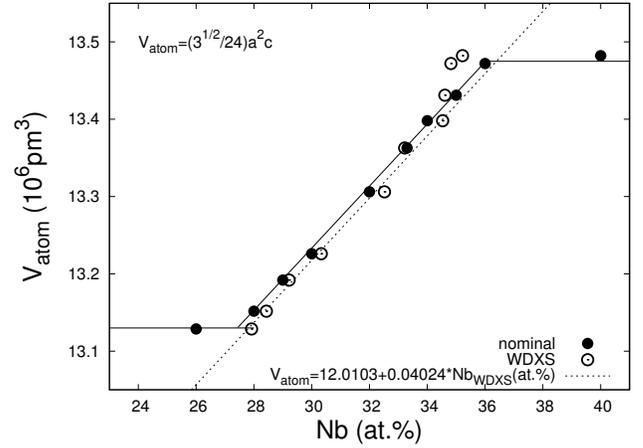}
\end{center}
\caption{Unit cell volume of the hexagonal C14 Laves phase \nfy calculated from the unit cell parameters, displayed in Table \ref{tab:table1}, versus the nominal and WDXS composition. The limits of the homogeneity region are 27.4 and 36.3 at.\% Nb. The dependence of the average atomic volume on the
composition determined by WDXS (dashed line) is used to extract the compositions of samples with small values of $y$.}
\label{fig:volume}
\end{figure}
According to EDXS analysis this phase is richer in Nb compared to the
Laves phase. The actual composition of this phase could not be accurately
determined due to its small grain size. It is plausible that in a sample
with a large amount of this Nb-rich second phase the C14
main phase will be Fe-rich. This argument may explain the difference in
lattice parameters between the samples with $y=0$ and $y=0.003$, which
were both extracted from the same ingot.

The main difference between the two methods emerged in the
metallographic analysis: While the first method resulted in high
homogeneity in all regions of the final pellet with very small
isolated filaments of the Ti$_{2}$Ni-like phase, the second method
produced drastic differences in homogeneity between the upper and the lower
zone inside the pellet. The zone with almost no extra phases appears
to be the middle of the pellet.

After having cut the samples into horizontal slices, we ground and polished them
carefully (in two steps) and observed the
surface with a high resolution optical microscope. The inset in
Fig.~\ref{fig:sample4b_p4} shows the optical micrograph taken from one of the upper
slices of the \nf polycrystal prepared with method B. It is the same
sample as that measured in Fig.~\ref{fig:x-ray}, where we observed the richest
amount of second phase. The phase can be seen clearly near the
C14-\nf grain boundaries, which have dimensions of about $50 \mu\rm m$.

Afterwards, we measured the AC-susceptibility at low magnetic fields
($H\leq 250 \rm{~Oe}$) of samples extracted from the upper and the
middle slice (Fig.~\ref{fig:sample4b_p4}): In the upper slice, we
observed a FM signal with a transition temperature $T_{s}\approx 80 \rm
~K$, in addition to the expected SDW peak at $T_{N}\approx 10 \rm
~K$. In the middle slice, only the SDW signature is present. Comparing
these measurements with the metallographic analysis, we conclude that
the second phase is responsible for the FM part of the
signal. Repeating the same procedure with different samples, we
observed that $T_{s}$ is different in every sample, varying from 250~K
to 15~K. It probably depends on the real Nb concentration at the grain
boundaries. Looking at the \nfy phase diagram
(Fig.~\ref{fig:phasediagram}), this hypothesis makes sense because the
Nb-rich region is FM and the transition temperature varies strongly
with the quantity of Nb. In magnetic fields higher than 250~Oe these
FM features disappear. It is therefore necessary to evaluate the
quality of any sample by not only measuring the residual resistance
ratio (RRR), but also the AC-susceptibility in zero magnetic field.

The lattice parameters of all polycrystals are listed in
Table~\ref{tab:table1}, together with the initial nominal
compositions, WDXS results and transition temperatures. The discrepancy
between the values of our lattice parameters for \nf and those
reported in literature is less than
0.001 \AA.\cite{shiga1987,bi1996,zhu2003} The measurements that we
are going to present have been performed on a selected number of
polycrystals, which show none or at most very low presence of the
second phase.

Fig.~\ref{fig:volume} shows the average volume per atom of the C14
structure versus the nominal (solid line) and WDXS (dashed line) Nb
content. The final composition of most samples, as determined by WDXS,
is slightly richer in Nb than the nominal composition. In the range from 28 to 34
at.$\%$ Nb, a linear dependence of the atomic volume on
composition determined by WDXS according to Vegard's volume is
observed. Around the upper border of the homogeneity range
($\approx35$~at.\% Nb), the sample composition could not be determined
accurately by WDXS, probably because of the large amount of the second
phase. Using the relation
\begin{equation}
V_{\mathrm{atom}}({\mathrm{\AA}}^3)=12.0103 +0.04024\cdot x ({\mathrm{at.\%~Nb_{WDXS}}})
\label{eq:volume}
\end{equation}
obtained by a least-squares fit of the data, we can determine
the composition of a sample by measuring the lattice parameters.
This is especially useful for detecting small
differences in composition that are difficult to observe using
WDXS. The accuracy in the WDXS experiments is of the order of $0.1\%$
limiting the accuracy on $y$ to $\sim 0.003$. We note that a small
systematic error of about 0.2 at.$\%$ cannot be ruled out. However,
this error would only shift the absolute value without changing the
shape of the phase diagram. 
\subsection{\label{subsec:experimental_setup}Experimental setup}
\noindent
Measurements of the resistivity, heat capacity, magnetization and
magnetic susceptibility down to 1.8~K have been carried out in a 9~T
Physical Properties Measurements System and in a 7~T Magnetic
Properties Measurements System (Quantum Design).
High-resolution measurements of the resistivity at temperatures down
to 50~mK were obtained in an adiabatic demagnetization refrigerator
(Cambridge Magnetic Refrigeration) by a standard 4-terminal lock-in
technique.
The low temperature heat capacity was determined in the same
refrigerator by a relaxation-time technique.
\begin{figure}
\begin{center}
\includegraphics[width=0.7\columnwidth,angle=-90]{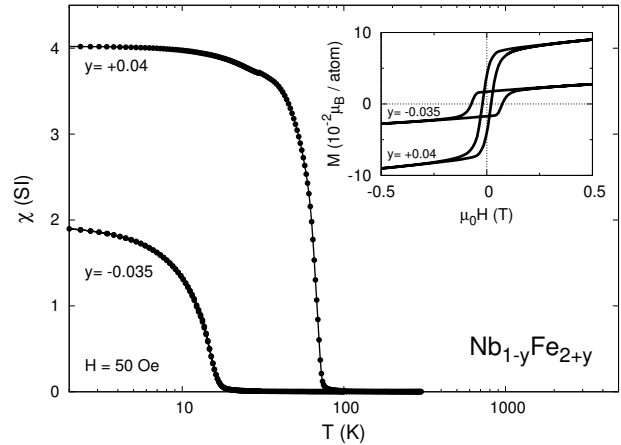}
\end{center}
\caption{Temperature dependence of the DC magnetic susceptibility of samples with $y=+0.04$ and $y=-0.035$.
Inset: Hysteretic magnetization curves for the same samples.
The Nb-rich sample shows a larger coercive field and smaller remanent magnetization.}
\label{fig:hysteresis}
\end{figure}
Due to the low molar specific heat of \nf compared to that of heavy
fermion compounds ($C_{p}\sim 48$ mJ/Kmol at 1~K), and due to the low sample
mass of $\sim 5 {\rm mg}$, we had to assemble a particular platform
for measuring the heat capacity with high resolution: We used sapphire
single-crystals of 6 mm diameter and $100 \mu$m thickness held
mechanically by four Pt$_{0.9}$Ir$_{0.1}$ wires, which provide a well
defined thermal link with a thermal conductance $K\sim0.2 \mu{\rm
  W/K}$ at 1~K; this configuration provides relaxation time constants
of about $\tau=C_{p}/K\sim 6$ s. Details of the heat capacity setup
will be described in a forthcoming article.
\section{\label{sec:results}Results and discussion}
\subsection{\label{sec:FMphases}FM phases: $\mid y\mid\geq 0.02$.}
\noindent
We focus our attention first on a very Nb-rich sample $(y=-0.035)$, as
well as on one very rich in Fe $(y=+0.04)$. For both compositions,
literature reports FM order.\cite{yamada1988,crook1995} They are
still in the C14 homogeneity range and the amount of the second phase
appears to be negligible. Our results confirm
that both Nb- and Fe-rich \nfy are FM, and indicate that the two
ferromagnetically ordered states are different in character.

Fig.~\ref{fig:hysteresis} shows the DC magnetic susceptibility of the
two samples measured at 50~Oe. Clear jumps in magnetization are
observed at $T\approx 72$~K and $T\approx 14.5$~K for the Fe-rich and
the Nb-rich sample respectively. When increasing the external field, the
transition temperature remains practically constant, suggesting that the
ferromagnetism is not a consequence of the second phase. In addition,
hysteresis signatures are evident in both samples, as shown in the
inset of the same figure. Although in both samples the remanent
magnetizations are very small, of the order of $10^{-2}\mu_{B}$ per
atom, the coercive fields are quite large: 200~Oe and 700~Oe.
\begin{figure}
\begin{center}
\includegraphics[width=0.9\columnwidth,angle=0]{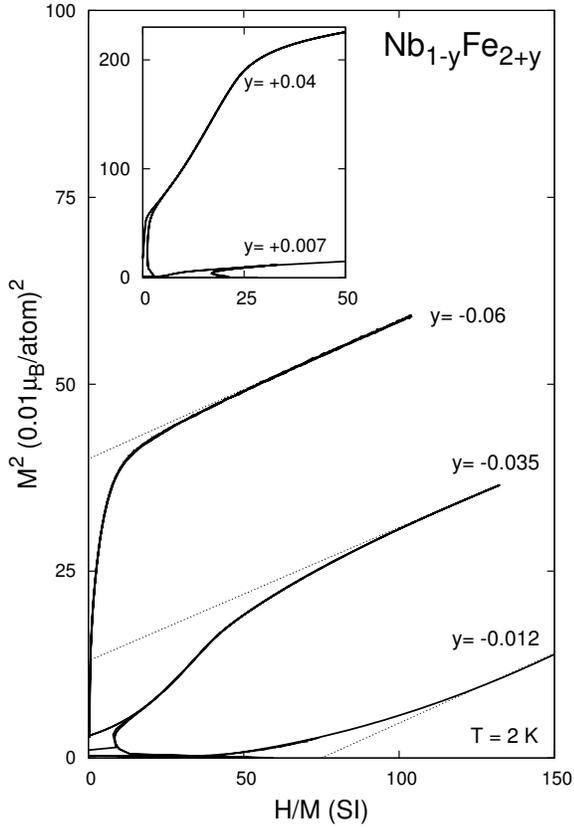}
\end{center}
\caption{Arrott plots for Nb-rich samples at 2~K. Inset: The same is for Fe-rich samples.}
\label{fig:arrott2K}
\end{figure}
%
With increasing Nb excess the coercive field of Nb-rich samples
decreases to a value that is not detectable anymore by the measurement
technique, as in the case of soft magnets: The Arrott plot for sample
$y=-0.06$ ($T_{c}=30$~K) in Fig.~\ref{fig:arrott2K} at 2~K does not
show any hysteresis, although the finite $M^2$ intercept for zero
$H/M$ is a clear indication of a remanent magnetization and hence of
ferromagnetism.  For very slight Nb doping, $y=-0.012$, no FM signal has
been observed down to 2~K and also the Arrott plot suggests a PM
groundstate.
\begin{figure}
\begin{center}
\includegraphics[width=0.7\columnwidth,angle=-90]{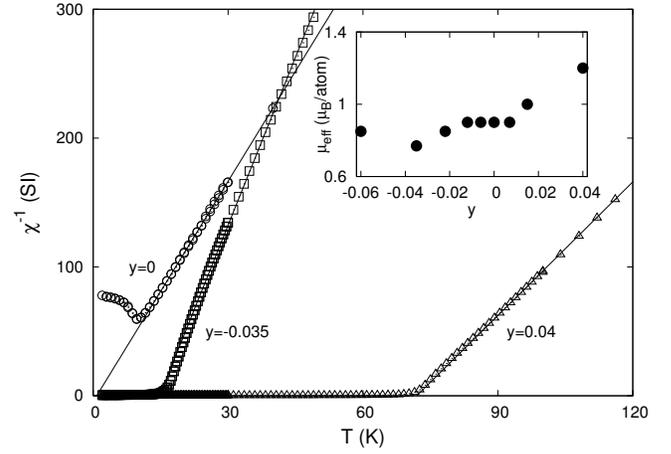}
\end{center}
\caption{Inverse of the magnetic susceptibility at three selected
  compositions versus temperature. All samples in the homogeneity
  range follow a Curie-Weiss law below 100~K. Inset: Dependence of the
  fluctuating moment $\mu_{eff}$ on $y$.}
\label{fig:curieweiss}
\end{figure}
On the Fe-rich side, all samples exhibited large hysteresis (see inset
of Fig.~\ref{fig:arrott2K}).  The FM transition in samples $y=+0.04$
and $y=-0.035$ is also inferred by the behavior of $\chi^{-1}$ versus
$T$ shown in Fig.~\ref{fig:curieweiss}. All samples in the homogeneity
range follow a Curie-Weiss law below 100~K with a slope $C=1/(T\chi)$
corresponding to a large fluctuating moment $\mu_{eff}$ compared to
the ordered moment ($\sim 0.02 \mu_{B}$), where
$C^{-1}=\frac{1}{3}\frac{N}{V}\mu_0\mu_{eff}^2/k_B$ and $\frac{N}{V}$ is
the atomic number density. 
%
\begin{figure}
\begin{center}
\includegraphics[width=0.7\columnwidth,angle=-90]{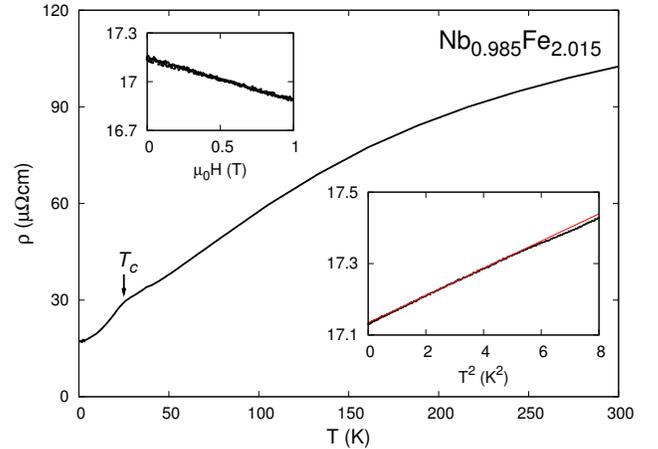}
\end{center}
\caption{(color online). Resistivity $\rho$ of sample with $y=+0.015$: The FM
  transition temperature $T_{c}$ is visible at 25~K where $\rho$
  suddenly decreases. Interestingly, $\rho$ follows a $T^{5/3}$ law
  above and below $T_{c}$ and recovers a Fermi-liquid $T^{2}$ dependence
  below 2~K (lower inset).\cite{brando2008} The magnetoresistance is
  almost flat at 100 mK (upper inset).}
\label{fig:rho7}
\end{figure}
As the composition is varied from slightly Fe-rich ($y>0$) to slightly
Nb-rich ($y<0$), $\mu_{eff}$ remains constant at $\simeq 0.9\mu_{B}$ per atom
(inset of Fig.~\ref{fig:curieweiss}), whereas the Curie-Weiss temperature
$\theta_{CW}$ changes sign, from
$\theta_{CW}\approx 3$~K for $y=0$ to $\theta_{CW}\approx -3$~K for
$y=-0.012$.

To study the ground state properties of the FM phase, we measured the
electrical resistivity and the heat capacity of a FM sample with
$y=+0.015$ down to 100~mK. Although the sample shows a small SDW
signal (peak in the magnetic susceptibility) at $T_{N}=32$~K, its
groundstate properties are dominated by the FM transition at
$T_{c}=25$~K: the resistivity $\rho(T)$, shown in Fig.~\ref{fig:rho7},
suddenly decreases only at the FM transition temperature $T_{c}=25$~K,
following a $T^{5/3}$ law above and below $T_{c}$
(Ref. 11),
which indicates a strong influence of FM
fluctuations on the electron scattering rate. Below $T_{FL}\simeq
2$~K, $\rho(T)$ recovers a Fermi-liquid $T^{2}$ dependence (lower
inset). The magnetoresistance is almost flat at 100~mK up to 1 Tesla
(upper inset).

\begin{figure}
\begin{center}
\includegraphics[width=0.7\columnwidth,angle=-90]{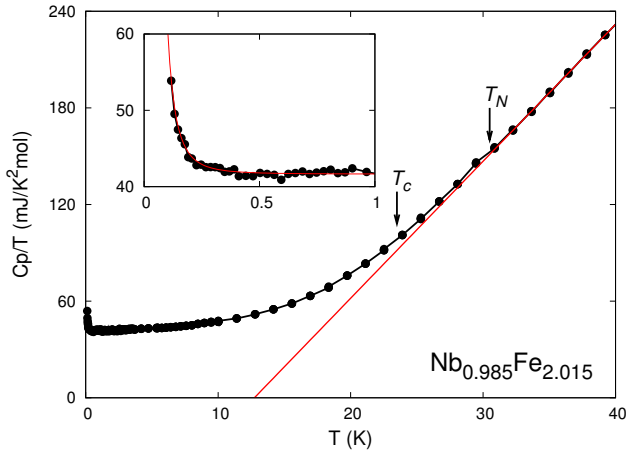}
\end{center}
\caption{(color online). Heat capacity divided by temperature $Cp/T$ vs. $T$ of the sample
  with $y=+0.015$: The SDW transition is slightly visible as a tiny
  hump at 32~K. Below 2~K $Cp/T$ is constant and increases
  dramatically below 300~mK, probably because of the nuclear Schottky
  contribution of the Nb atom surrounded by a strong FM field
  (inset). The line in the inset is a fit with a Schottky function
  $Cp/T \sim T^{-3}$.}
\label{fig:CpTvsTsample7}
\end{figure}

The temperature dependence of the heat capacity divided by
temperature, $C_p/T$, of the sample with $y=+0.015$ is shown in
Fig.~\ref{fig:CpTvsTsample7}. No clear heat capacity anomalies are
visible at the SDW and FM transition temperatures (arrows), which
suggests small ordered moments. Below 2~K, $C_p/T$ levels off at a
value of 42 mJ/K$^{2}$mol, consistent with a Fermi-liquid ground state.

Below 300~mK, however, $C_p/T$ increases dramatically, probably
because of the nuclear Schottky contribution of the Nb atoms subject
to a the internal field of a ferromagnet (inset). By fitting the points below
2~K with a Schottky contribution, we obtain $C_p/T=(41.6+0.02 (T/{\rm
K})^{-3}) {\rm mJ/K ^{2} mol}$. From this equation we can estimate
the internal magnetic field $B_{eff}$ experienced by the Nb atoms:
\begin{displaymath}
  0.02 {\rm K mJ/mol}=\frac{R}{3}I(I+1)\left(\frac{h\gamma_{NMR}}{k_{B}}\right)^{2}B^{2}_{eff}
\end{displaymath}
where $\gamma_{NMR}$ is given by NMR experiments and $I$ is the
nuclear spin of the Nb atoms. For $^{93}$Nb atoms with $I=9/2$ and
$\gamma_{NMR}=10.4$ MHz/T, the resulting field is
$B_{eff}=1.08$~T. Only in this FM sample (the only FM sample measured
down to 50~mK), has the Schottky contribution to the heat capacity
been clearly observed.

We conclude that the groundstate of the Fe-rich compounds is
ferromagnetic and that the low temperature transport and thermodynamic
properties are well explained by the Fermi-liquid theory, as in many
other FM transition metal compounds.\cite{niklowitz2005}
\subsection{\label{subsec:SDWphase}SDW phase: $y = 0$}
We focus now on stoichiometric NbFe$_{2}$. Based on the broadening of the NMR
line width and on the ``S''-like feature in the magnetization curves
below $T_{N}=10$~K, \nf had been reported to exhibit spin density wave
order at low temperature.\cite{shiga1987,yamada1988,crook1995} This
interpretation is consistent with our measurements of the magnetic
susceptibility, specific heat capacity and
magnetization.\cite{brando2008,moroni2005,brando2006} Based on these
earlier results, we suggested that the low temperature magnetic order
in stoichiometric \nf may take the form of a SDW
with an helical arrangement of the Fe spins. Direct evidence for SDW
order in \nf from neutron scattering is still
outstanding,\cite{cywinski} but a renewed attempt, using large single
crystals grown in an infrared mirror furnace, is scheduled.
\begin{figure}
\begin{center}
\includegraphics[width=0.85\columnwidth,angle=0]{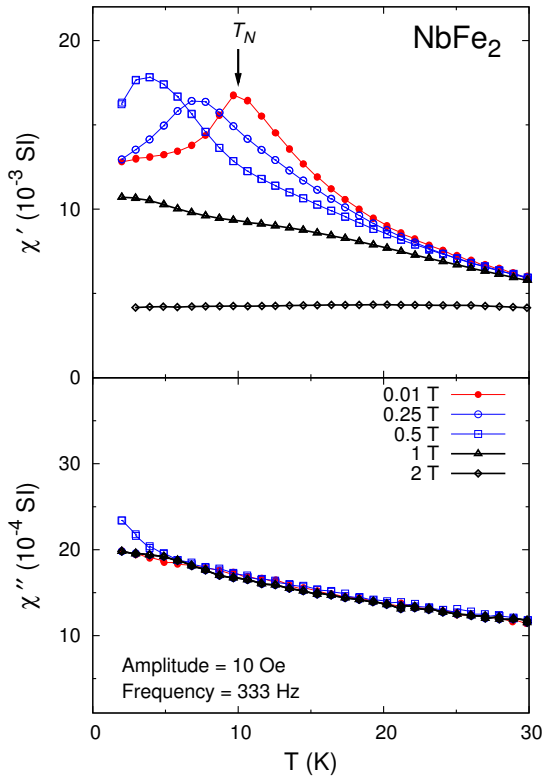}
\end{center}
\caption{(color online). Real and imaginary part of the AC-susceptibility in different magnetic fields for NbFe$_{2}$.}
\label{fig:chi12vsTsample4}
\end{figure}
%

Fig.~\ref{fig:chi12vsTsample4} shows the real and imaginary
components of the AC-susceptibility. The distinct SDW peak
is visible at 10~K, cutting off the Curie-Weiss-like behavior with
decreasing temperature (see Fig.~\ref{fig:curieweiss}). The low
transition temperature, for a \textit{d}-metal compound, as well as the
geometrically frustrated Kagom\'e structure of the Fe atoms in the C14
phase, might suggest a spin-glass transition.
\begin{figure}
\begin{center}
\includegraphics[width=0.85\columnwidth,angle=0]{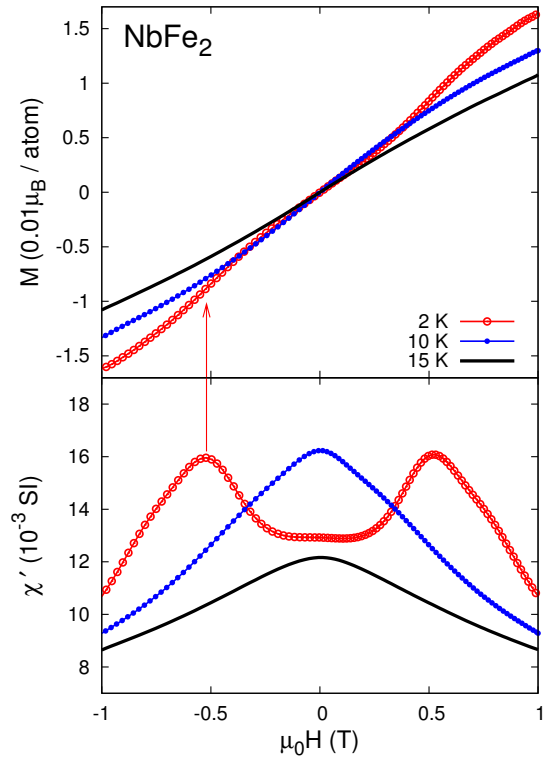}
\end{center}
\caption{(color online). Magnetization (upper panel) and (lower panel) AC-susceptibility vs. field $H$.}
\label{fig:MvsBsample4}
\end{figure}
However, the measured Curie-Weiss temperature $\theta_{CW}$ is close
to 1~K giving a frustration factor $f=\theta_{CW}/T_{N}$ of only 0.1.
By contrast, the hall-mark of frustration in local moment systems is a
frustration factor $f>1$. Moreover, the lack of a loss peak in
$\chi''$ at $T_{N}$, as well as the absence of a distinct frequency
dependence of $\chi'$ argue against a spin-glass transition.

The analysis of the irreversible field-cooled (FC) and
zero-field-cooled (ZFC) measurements (not shown) indicates a slight
splitting of the magnetization curves at $T_{N}$, but the effect is so
small that it would be difficult to attribute it to the main
phase. Only at much lower temperatures ($T\leq 3$~K) and at very low
fields does this splitting become more pronounced, suggesting the
formation of magnetic domains.\cite{dennis-thesis2006} This behavior
will be analysed in the next section, where it resurfaces in a
slightly Fe-doped sample.

\begin{figure}
\begin{center}
\includegraphics[width=\columnwidth,angle=0]{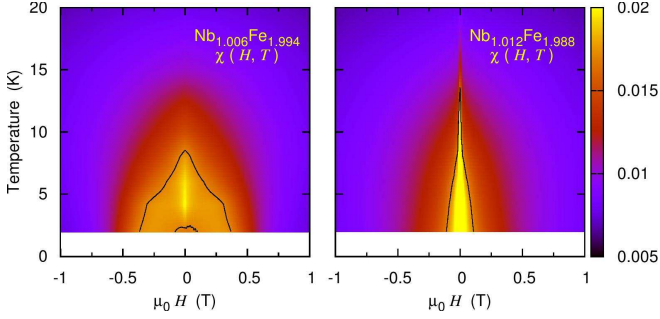}
\end{center}
\caption{(color online). Color plots of the magnetic susceptibility for Nb-rich samples at 2~K. The SDW phase in $y=-0.006$, defined by the bright zone, shrinks to a point for $y=-0.012$.}
\label{fig:Chi3D}
\end{figure}

An applied magnetic field $H$ shifts the $\chi'$ peak to lower $T$ and
suppresses the magnetic order at a critical field $\mu_{0}H_{c}\approx
0.6 \rm ~T$ at $100 {\rm mK}$, switching the system to a paramagnetic
state.  This can be seen more clearly in the magnetization and
susceptibility plots in Fig.~\ref{fig:MvsBsample4}. Such a behavior
could be interpreted as a metamagnetic transition from a paramagnetic
(PM) state to another PM state with a higher magnetization in
field. Such transitions are not uncommon in nearly FM metals, such as
YCo$_{2}$,\cite{sakakibara1990} but here, in contrast to the
metamagnetic PM-to-PM case, $H_{c}$ is shifted to lower values with
increasing temperature (see color plot of Fig.~\ref{fig:Chi3D}).\cite{brando2008}
\begin{figure}
\begin{center}
\includegraphics[width=0.85\columnwidth,angle=0]{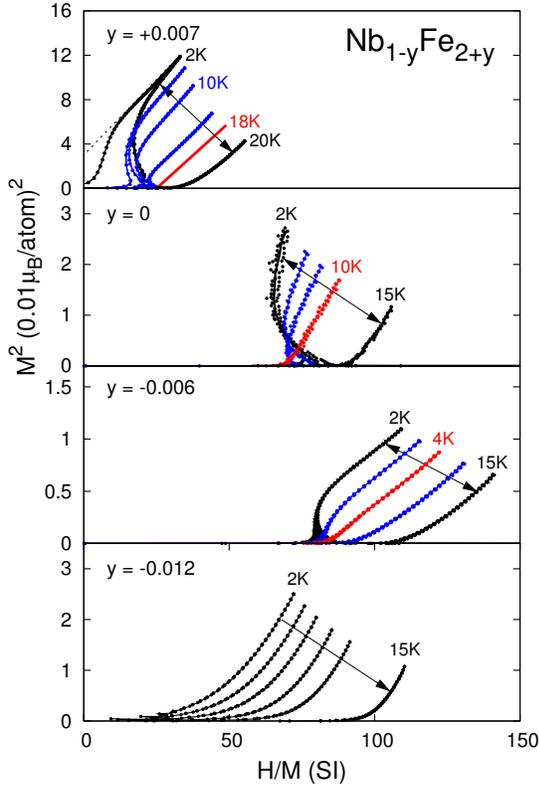}
\end{center}
\caption{(color online). Arrott plots for four samples with $y$ varying from small positive to negative values. The $M^{2}$ vs. $H/M$ lines are changing slope exactly at $T_{N}$ where the extrapolation of the linear dependency crosses the $y$ axis at negative value of $M^{2}$.}
\label{fig:Arrotty}
\end{figure}

\begin{figure}
\begin{center}
\includegraphics[width=0.7\columnwidth,angle=-90]{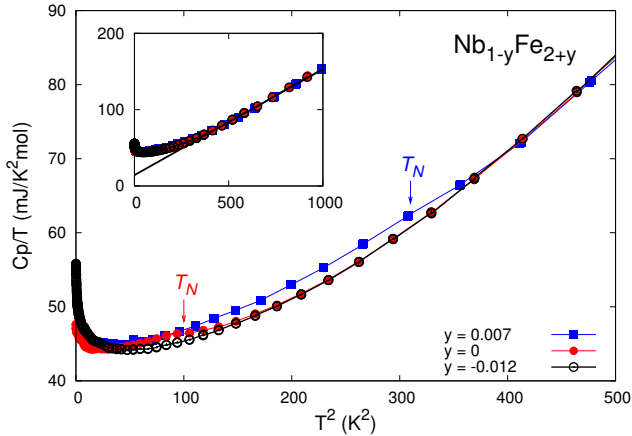}
\end{center}
\caption{(color online). $Cp/T$ vs. $T^{2}$ of two SDW samples with $y=0.007$ and
  $y=0$ and a paramagnetic one with $y=-0.012$. The inset shows the
  high temperature data fitted with a linear function
  (Eq.~\ref{eq:CpT}).}
\label{fig:CpTvsT2}
\end{figure}

A further indication of a bulk SDW transition at $T_N$ is given by the
Arrott plot in Fig.~\ref{fig:Arrotty}, in which the high field,
high temperature linear dependence of $M^2$ on $H/M$ changes over to
an arc with negative slope below $T_N$ and $H_c$.\cite{yamada1988}
Moreover, in the specific heat capacity (Fig.~\ref{fig:CpTvsT2}), the
bulk SDW transition is visible as a small hump at $T_{N}$, when the
data are compared with measurements in magnetic fields $H\geq
H_{c}$.\cite{brando2006} In Fig.~\ref{fig:CpTvsT2}, this hump is shown
not only for $y=0$ but also for the slightly Fe-rich sample with
$y=0.007$, which shows signatures of an SDW transition as well, at
$T_N \simeq 18 {\rm ~K}$.
The entropy below the transition is very low, of the order of $4 {\rm
  mJ/Kmol}$. Fig.~\ref{fig:CpTvsT2} also shows the heat capacity of a
slightly Nb-rich sample ($y=-0.012$). At this composition, no
indication of a transition has been observed in the heat capacity down
to 2~K. The inset of the Fig.~\ref{fig:CpTvsT2} shows how the high
temperature data can be fitted with the electronic and phononic
contribution:
\begin{equation}
Cp/T=\gamma_{el}+5832 {\rm J/K mol} \left(\frac{T^{2}}{\theta_{D}^3}\right)
\label{eq:CpT}
\end{equation}
where the Sommerfeld coefficient is $\gamma_{el}=14.2 {\rm mJ/K^{2} mol}$
and the Debye temperature is $\theta_{D}=348 \rm ~K$. This Sommerfeld
coefficient agrees very well with that predicted by band-structure
calculations.\cite{takayama1988,inoue1989} However, $Cp/T$ increases
towards low temperatures $T< 20 \rm ~K$ and saturates below 1~K near a
value of $\approx 48 {\rm mJ/K^{2}mol}$ (cf. Fig.~\ref{fig:CpTvslogT}),\cite{wada1990}
three times larger than the band structure value.
%
\begin{figure}
\begin{center}
\includegraphics[width=0.85\columnwidth,angle=0]{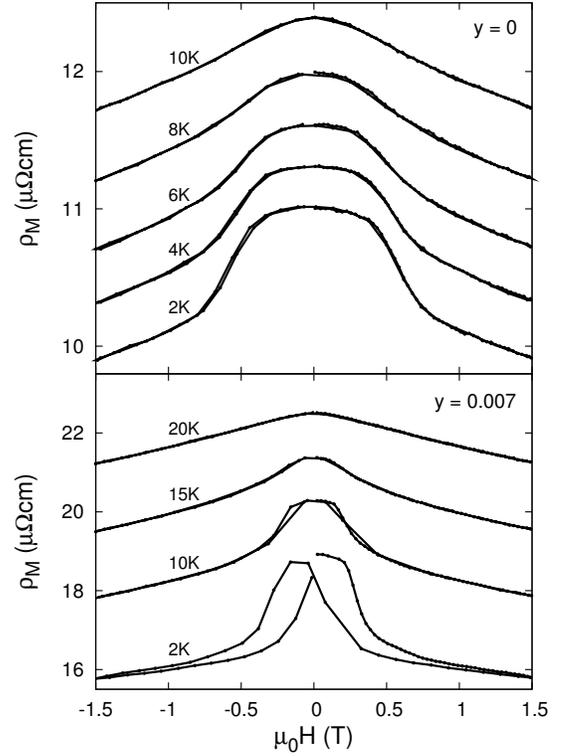}
\end{center}
\caption{Magnetoresistance of the stoichiometric sample (upper panel) and the slightly Fe-rich sample (lower panel) at different temperatures.}
\label{fig:rhovsB}
\end{figure}
The most dramatic signature of the SDW state is the behavior of
the magnetoresistance $\rho_{M}$: Although the SDW transition cannot
be detected in the temperature dependence of the resistivity, the low
temperature magnetoresistance $\rho_{M}$ shows a jump of about $10\%$
at the critical field in NbFe$_{2}$, and even more ($\approx 17\%$) in the
Fe-rich sample with $y=0.007$, as shown in Fig.~\ref{fig:rhovsB}. The
magnitude of the jump thus appears to be related to $T_{N}$. At $T_{N}$
this feature disappears and $\rho_{M}$ follows a linear field
dependence at high field, as in a paramagnetic metal.

In NbFe$_{2}$, all magnetic signatures below $T_{N}$ are practically identical
to those observed in the helically ordered compound MnAu$_{2}$,
although the latter has a much higher transition temperature
$T_{N}\simeq 380 \rm ~K$.\cite{samata1998} Furthermore, the large
magnetic anisotropy observed in single crystals,\cite{kurisu1997} and
the the critical field $\mu_{0}H_{c}=0.6 \rm ~T$ required to destroy the SDW
state, which is a surprisingly low value when compared to $T_{N}$,
suggest a long wavelength incommensurate magnetic
groundstate. Moreover, the hexagonal C14 structure may favor helical
order with propagation wavevector along the $c$-axis, as in rare earth
metals.\cite{blundell2001} Finally, the large Stoner factor $S\simeq
180$ ($I\cdot N(\epsilon_{F})=0.99\approx 1$), computed by comparing
the value of $\chi$ at $T_{N}$ to the one expected from band-structure
DOS,\cite{takayama1988,inoue1989} and the high Wilson ratio of about
60,\cite{brando2008} indicates that this compound is on the border of
a ferromagnetic instability. All these observations led us to propose
that the SDW state grows out of the $Q=0$ ferromagnetism existing in
Fe-rich NbFe$_{2}$, and develops into a $Q \neq 0$ helical state (as illustrated
in Fig.~\ref{fig:FMAFM3a}).
\subsection{\label{subsec:mixedphase}SDW and FM mixed phase: $0\leq y\leq 0.02$}
\begin{figure}
\begin{center}
\includegraphics[width=0.7\columnwidth,angle=-90]{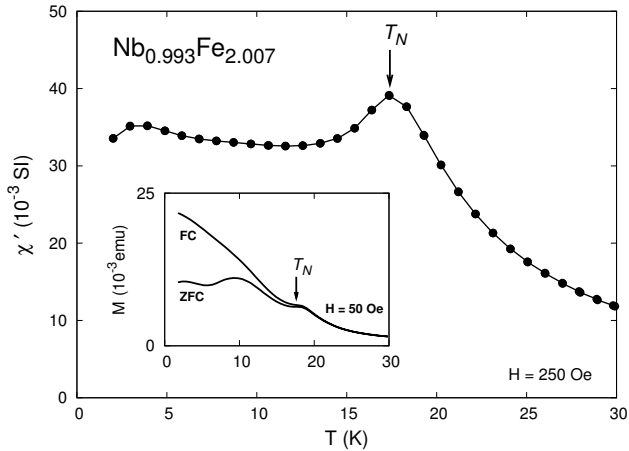}
\end{center}
\caption{AC-susceptibility for $y=0.007$ in a field of 0.025 T. The inset shows the ZFC-FC measurement performed with 50~Oe.}
\label{fig:chi12vsTsample325}
\end{figure}

\noindent
To examine our hypothesis that the SDW state grows continuously out of
the FM phase in Fe-rich samples, we measured a slightly Fe-doped
sample with $y=0.007$, which is located in the phase diagram around
the point of crossover between the pure FM phase and the SDW one. The
AC-susceptibility, plotted in Fig.~\ref{fig:chi12vsTsample325}, shows
the same sharp peak as in NbFe$_{2}$, but at a higher temperature
$T_{N}=18 \rm ~K$. The bulk transition is also visible in the specific heat
measurement shown in Fig.~\ref{fig:CpTvsT2}.
The ZFC-FC curves separate exactly at $T_{N}$, whereas in the
stoichiometric ($y=0$) sample a clear separation only occurred well
below $T_N$.\cite{sato-turtelli2007} This might indicate spin glass behavior.
However, as in the stoichiometric sample, we did not observe any
shift of the $\chi'$ maximum at $T_{N}$ with measurement frequency.

Hysteresis phenomena appear only at temperatures below $10 \rm ~K$, as
shown, for example, in the Arrott plot of Fig.~\ref{fig:arrott2K} at
2~K: In this plot the sample seems to be SDW between 10~K and 18~K, but
develops a remanent magnetization suggesting FM at lower
temperatures. These features are illustrated more drastically in the $\rho_{M}$
isotherms of Fig.~\ref{fig:rhovsB}: Below $T_{N}$, the sample develops
a clear negative magnetoresistance, and $\rho_M$ jumps towards lower
values as the critical field $H_c$ is exceeded.  The magnitude of the jump
increases with decreasing temperature. Around $10 \rm ~K$, $\rho_{M}$ begins
to shows hysteresis until at 2~K the ``up'' and ``down''
lines are well separated. The resulting curves suggest that the
hysteresis phenomena detected in the stoichiometric samples are not a
consequence of secondary FM phases, but are an intrinsic property of the
SDW state.

\begin{figure}
\begin{center}
\includegraphics[width=0.85\columnwidth,angle=0]{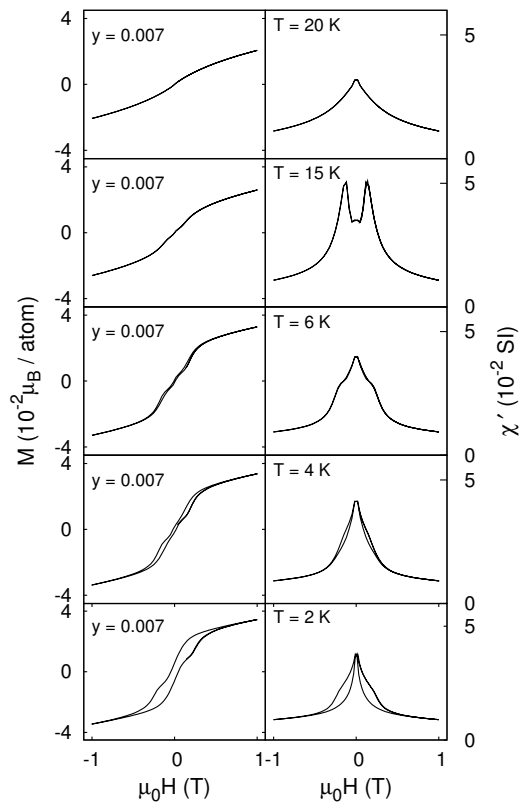}
\end{center}
\caption{Sequence of isothermals curves (magnetization and AC-susceptibility) taken at temperatures above $T_{N}=18$~K and far below. This sequence shows how with decreasing temperature the hysteresis opens first at the critical field $\mu_{0}H_{c}=0.15$ ~T and then becomes broad also at $H=0$.}
\label{fig:sequence325}
\end{figure}

Supporting evidence for this interpretation is provided in
Fig.~\ref{fig:sequence325}, which shows a sequence of magnetization
and AC-susceptibility isotherms for the same sample.  As already
pointed out by Crook and Cywinski,\cite{crook1995} the hysteresis
lines separate firstly not at $H=0$ but at finite fields. Following
the measurements from the top to the bottom frame of the figure, we
recognize the double peak at $15 \rm ~K$ and $\mu_{0}H_{c}=0.15 \rm ~T$ as a
signature of the SDW phase. As the temperature is lowered further,
between $15 \rm ~K$ and $4 \rm ~K$ hysteresis develops, which straddles
$H_c$ but does not yet reach $H=0$. Finally, close to $2 \rm ~K$, the
size of the hysteretic field region includes zero field, so that the
two hysteresis regions associated with $+H_c$ and $-H_c$ merge, giving
rise to a FM remanent magnetization.
All investigated samples with Fe concentrations $0\leq y \leq 0.02$
display qualitatively the same phenomena, only the transition
temperatures and the magnitudes of the effects change with
increasing $y$. For $y \geq 0.02$ the samples seem to show only the FM
character. The fact that in these very iron-rich, purely FM samples
the magnetoresistance $\rho_{M}$ is much weaker and does not show any hysteresis
suggests that the hysteretic signatures in $M(H)$,
$\rho_{M}(H)$ and $\chi'(H)$ are a characteristic property of this
specific SDW state and indicates the presence of SDW domains, as
recently discussed by Jaramillo\et.\cite{jaramillo2007}
\subsection{\label{subsec:PMphase}PM phase: $y\approx -0.015$}
\noindent
While Fe-excess increases the SDW ordering temperature $T_N$, and
eventually induces low-ordered-moment ferromagnetism, slight Nb-excess
weakens the SDW state and tunes the system towards paramagnetism
and therefore to a QCP. This has been observed in the behavior of the
magnetic susceptibility $\chi$ in two samples with $y=-0.006$ and
$y=-0.012$. Fig.~\ref{fig:Chi3D} shows the color plots of $\chi(H,T)$
for the two samples at $2 \rm ~K$: The ordered SDW phase for the $y=-0.006$
sample, defined by the bright zone in the $H-T$ diagram, shrinks to a
point for $y=-0.012$.
A similar plot for a stoichiometric sample ($y=0$) is shown in
Ref.~11.
%
\begin{figure}
\begin{center}
\includegraphics[width=0.7\columnwidth,angle=-90]{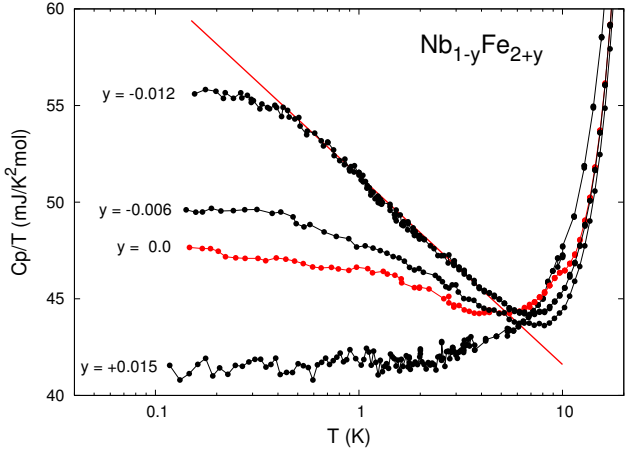}
\end{center}
\caption{(color online). Specific heat of samples with $y$ close to 0. The low temperature Schottky contribution for the sample with $y=0.015$ has been substracted.}
\label{fig:CpTvslogT}
\end{figure}

Additional evidence for this evolution towards a
paramagnetic state close to $y=-0.015$ is given by the sequence
of Arrott plots shown in Fig.~\ref{fig:Arrotty}. For $y=0$ and
$y=-0.006$, the $M^{2}$ vs. $H/M$ curves change appearance at $T_N$:
for $T>T_N$, the curves bend towards the left, whereas for $T<T_N$
they bend towards the right. By contrast, for $y=-0.012$, no such
change in behavior occurs down to $2 \rm ~K$.

The heat capacity is less sensitive to the magnetic transition and
so the weak anomaly that was observed in the stoichiometric sample
at 10~K, has not yet been observed in either Nb-rich sample.
The surprisingly high value of
the Sommerfeld coefficient $\gamma$ at low temperature for all the
samples close to the stoichiometric composition has been considered as
indication of the presence of strong spin fluctuations and of the
vicinity of the system to a QCP.\cite{brando2006,brando2007}
Approaching the QCP by increasing the Nb-content, $\gamma$ gradually
increases and, for $y=-0.012$, it follows a logarithmic temperature
dependence down to about $0.6 \rm ~K$ (Fig.~\ref{fig:CpTvslogT}).

\begin{figure}
\begin{center}
\includegraphics[width=0.7\columnwidth,angle=-90]{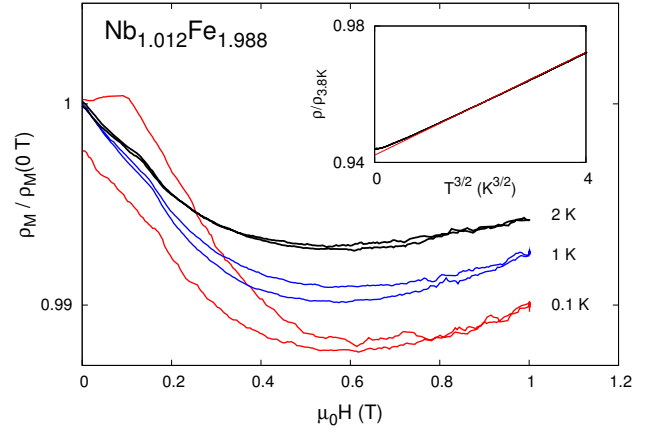}
\end{center}
\caption{(color online). Magnetoresistance of the sample with $y=-0.012$ at three
  temperatures. Inset: $T^{3/2}$ dependence of the resistivity; below
  0.6~K the measured curve deviates strongly from the linear fit.}
\label{fig:res_676}
\end{figure}

Non-Fermi-liquid behavior has also been observed in the temperature
dependence of the electrical resistivity. The inset of
Fig.~\ref{fig:res_676} shows the temperature dependence of the
resistivity $\rho(T)$: its power-law exponent
$n=d(log\Delta\rho)/d(logT)$ varies from $3/2$ to $5/3$ between $3 \rm
~K$ and $0.6 \rm ~K$ and crosses over to the FL exponent 2 below
$0.6 \rm ~K$.\cite{brando2007} Combining both observations,
the $T$-dependence of the heat capacity and of the resistivity below 3~K,
the sample Nb$_{1.012}$Fe$_{1.988}$ seems to be close to
a ferromagnetic QCP, where $C/T \propto log(T)$ and $\rho \propto
T^{5/3}$ have been predicted for a 3D FM metal.\cite{moriya1985,lonzarich1997}
Below $0.6 \rm ~K$, Nb$_{1.012}$Fe$_{1.988}$ experiences a crossover to the
conventional FL state.

%

Magnetoresistance measurements below $2 \rm ~K$
(Fig.~\ref{fig:res_676}) suggest that \nfy does not have a
paramagnetic ground state at $y=-0.012$. A little drop (less than
1$\%$) in $\rho_{M}$ is observed at 2~K and clear hysteresis
phenomena develop below $1 \rm ~K$. This indicates the presence of very
weak SDW order below $T_N \simeq 2 \rm ~K$, which develops domains at
$T\leq 1 \rm ~K$.
%

These findings contrast with our observations on a high quality
Nb-rich ($y\simeq -0.01$) single crystal, in which non-Fermi liquid
phenomena have been observed down to the lowest temperature.\cite{brando2008}
Despite residual low-temperature order below
$T_N\simeq 2.8 \rm ~K$, presumably with a very small ordered moment,
the resistance of the single crystal follows a $T^{3/2}$ power law
down to $50{\rm ~mK}$, and its heat capacity coefficient $C_{p}/T$
increases logarithmically with decreasing temperature, without any
indication of a crossover to a FL state. The return of our Nb-rich
polycrystal with $y=-0.012$ to FL behavior at low temperature, when
such a return to FL behavior is not observed in a single crystal with
a similar composition, may be attributed to the lower quality of
the polycrystal either in terms of its purity or its homogeneity. 

We interpret these findings as the first clear indication of
a logarithmic breakdown of the Fermi-liquid state in a transition
metal antiferromagnet: The fact that the precise power-law exponent
varies between 3/2 and 5/3 and the discrepancies between the phenomena
observed in single and polycrystals suggest that the precise form of
$\rho(T)$ may depend on stoichiometry and on the sample orientation
and that present data cannot decide whether FM or SDW spin
fluctuations determine the low temperature behavior.
\section{\label{sec:phasediagram}Phase diagram and conclusion}
\begin{figure}
\begin{center}
\includegraphics[width=0.7\columnwidth,angle=-90]{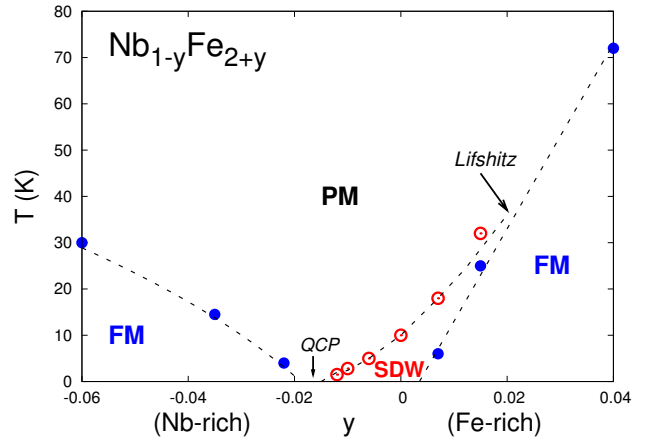}
\end{center}
\caption{(color online). Proposed new phase diagram for Nb$_{1-y}$Fe$_{2+y}$.}
\label{fig:phasediagram}
\end{figure}

\noindent
We propose a refined version of the magnetic phase diagram for \nfy
(Fig.~\ref{fig:phasediagram}), obtained by combining all the
thermodynamic and transport measurements discussed in this paper. 
Dashed lines indicate the likely limits of the FM and SDW state.
We can address the five questions outlined in the
introduction with the help of this phase diagram.

For $\mid y \mid \geq 0.02$, distinct FM phases have been observed at
low temperature. Our data confirm that the FM features do not originate from
an impurity phase, but are intrinsic to the C14 Laves phase Nb$_{1-y}$Fe$_{2+y}$.
The $M-H$ isotherms are qualitatively different in the Fe-rich and Nb-rich
FM phases, suggesting that the two states are of a different nature.
%

\begin{figure}
\begin{center}
\includegraphics[width=0.6\columnwidth,angle=-90]{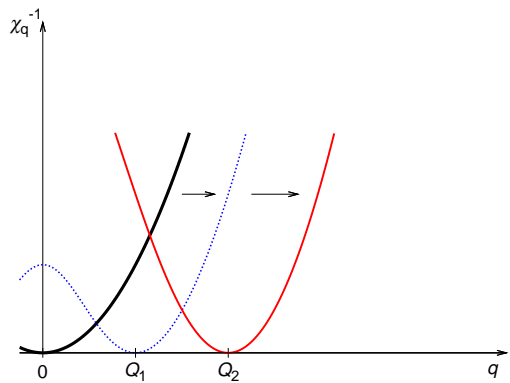}
\end{center}
\caption{(color online). Postulated evolution of the inverse of the wavevector $q$
  dependent susceptibility $\chi_q$ with doping. With decreasing iron
  content $y$ the minimum of the dispersion $\chi_q^{-1}$ may shift
  from $q=0$ to finite $q$. This shift is accompanied by a change from
  FM to SDW, possibly helical, magnetic order. The point in the phase
  diagram, where this crossover takes place, is called Lifshitz point
  (Fig.~\ref{fig:phasediagram}).}
\label{fig:FMAFM3a}
\end{figure}

Between the FM phases, a distinct non-FM phase has been observed:
susceptibility, magnetization, resistivity and heat capacity results,
many of which are similar to observations in well-known spin helical
compounds such as MnAu$_2$,\cite{samata1998} Y$_2$Fe$_{17}$\cite{prokhnenko2005}
and MnSi,\cite{pfleiderer1997,stishov2007} suggest that the
magnetic order in stoichiometric \nf assumes the form of a SDW,
possibly a long wavelength helical or spiral state.
On the Fe-rich side, we believe that the SDW
state with a small wave vector $Q \neq 0$ connects continuously to the
FM state at the Lifshitz point, which is located
around $y=+0.02$.\cite{hornreich1975} The schematic
Fig.~\ref{fig:FMAFM3a} shows the possible evolution of the wave vector
dependent generalised susceptibility $\chi(q)$, as \nfy is tuned away
from the Lifshitz point: With decreasing iron content $y$ the minimum
shifts from $q=0$ to low $q \rightarrow Q_{1} \rightarrow Q_{2}$, and
\nfy changes from FM to long-wavelength SDW order. The small value of
the $\chi({q})^{-1}$ intercept at $q=0$ would explain the high value
of the measured AC-susceptibility at low temperature. Assuming that the
dispersion stiffness value of \nf is similar to the one measured
in ZrZn$_{2}$, we could estimate an ordering wave vector of about
0.05 \AA$^{-1}$ for NbFe$_{2}$.\cite{brando2008} Until the precise nature of
the SDW state can be observed using a microscopic probe, ideally
neutron scattering, this scenario remains strongly
hypothetical.
Alternative origins of the dual nature of \nf -- nearly
ferromagnetic and yet antiferromagnetic -- include (i) coincidental
peaks of the wave vector dependent susceptibility at large wave vector
(SDW) and at $q=0$ (FM) and (ii) a very weakly dispersive wave vector
dependent susceptibility $\chi(q)$, peaking at a large,
antiferromagnetic wave vector but remaining unusually large even at
$q=0$. Neutron scattering experiments on recently grown, large single
crystals of \nf are planned, and will hopefully help to resolve this
controversy.

Once the SDW state is entered, SDW domains form with different order
parameter polarizations or ordering wavevector directions, creating a
set of domain walls. A weak magnetic field $H$ is sufficient to
destroy the SDW and induce a weakly spin-aligned paramagnetic
state. The transition in $H$ from helical to the paramagnetic state
probably occurs through a number of successive steps, which are linked
with domain wall jumps and define the size of the hysteresis. We are
planning to investigate this interpretation by high precision
measurements of the magnetic field dependence of the magnetization
and of the electrical resistivity at very low temperatures.
By reducing the Fe content (for small negative $y$), the SDW state can
be suppressed completely, leading to a magnetic quantum critical point
and leaving a small paramagnetic region around $y\approx-0.015$. Here,
magnetic fluctuations dominate the temperature dependence of
resistivity and specific heat. Since the critical exponents fall in
between those expected for FM and SDW spin fluctuations and the QCP is
located between the two different phases, the nature of the
fluctuations requires clarification.
%
\begin{acknowledgments}
We are indebted to R. Ballou, A. Chubukov, B. F\aa k, S. Grigera, G. G. Lonzarich, P. Niklowitz, C. Pleiderer, J. Quintanilla, J. Saunders, A. Schofield, B. Simons and Y. Yamada for useful discussions. Part of this work has been founded by UK Government's funding agency EPSRC, grant number GR/T09866/01. Moreover, we thank the Max-Planck Inter-Institutional Research Initiative ``The Nature of Laves Phases'' and A. Kerkau for supporting key characterization activities.
\end{acknowledgments}
\end{document}